\documentclass{aa}  

\usepackage{comment}
\usepackage{hyperref}
\usepackage{csquotes}
\usepackage{graphicx}
\usepackage{xcolor}
\usepackage{placeins}
\usepackage{txfonts}
\usepackage{outlines}
\usepackage[separate-uncertainty=true,multi-part-units=single]{siunitx}

\DeclareSIUnit{\erg}{erg}
\DeclareSIUnit{\parsec}{pc}
\DeclareSIUnit{\jansky}{Jy}
\DeclareSIUnit{\beam}{beam}

\begin{document}

   \title{Serendipitous decametre detection of ultra steep spectrum radio emission in Abell\,655}

   \subtitle{}

   \author{C. Groeneveld \inst{\ref{inst1}} \and
   R.~J. van Weeren \inst{\ref{inst1}} \and
   A. Botteon \inst{\ref{inst2}} \and
   R. Cassano \inst{\ref{inst2}} \and
   F. de Gasperin \inst{\ref{inst2}} \and
   E. Osinga \inst{\ref{inst3} } \and
   G. Brunetti \inst{\ref{inst2}} \and
   H.~J.~A. R\"ottgering \inst{\ref{inst1}}}

   \institute{Leiden Observatory, Leiden University, Einsteinweg 55, 2333 CC Leiden, the Netherlands \label{inst1} \and
   INAF-Istituto di Radioastronomia, via Gobetti 101, 40129 Bologna, Italy \label{inst2}\and
   Dunlap Institute for Astronomy and Astrophysics, University of Toronto, 50 St. George St, Toronto, ON M5S 3H4, Canada \label{inst3}}

   \date{Received XXXX, Accepted YYYY}

  \abstract{Some galaxy clusters contain non-thermal synchrotron emitting plasma that permeate the intracluster medium (ICM). The spectral properties of this radio emission are not well characterised at decametre wavelengths ($\nu < \SI{30}{\mega\hertz}$), primarily due to the severe corrupting effects of the ionosphere. Using a recently developed calibration strategy, we present LOFAR images below $\SI{30}{\mega\hertz}$ of the low-mass galaxy cluster Abell\,655, which was serendipitously detected in an observation of the bright calibrator 3C\,196. We combine this observation with LOFAR data at 144\,MHz and new band 4 upgraded Giant Metrewave Radio Telescope observations centred at 650\,MHz.
  In the 15--30\, MHz LOFAR image, diffuse emission is seen with a physical extent of about 700~kpc.  
  We argue that the diffuse emission detected in this galaxy cluster likely has multiple origins. At higher frequencies ($\SI{650}{\mega\hertz}$), the diffuse emission resembles a radio halo, while at lower frequencies the emission seems to consist of several components and bar-like structures. 
  This detection of diffuse emission suggests that most low-frequency emission in this cluster comes from re-energised fossil plasma from old Active Galacitic Nucleus outbursts co-existing with the radio halo component.
  By counting the number of cluster radio detections in the decametre band, we estimate that around a quarter of the Planck clusters host re-energised fossil plasma that is detectable in the decametre band with LOFAR.}

   \keywords{Galaxies: clusters: individual -- Radio Continuum: general -- Techniques: Interferometric }

   \maketitle

\section{Introduction}
\label{sec:intro}

Galaxy clusters are the most massive virialised structures in the Universe.
The majority of the baryonic matter in clusters is contained in the intracluster medium (ICM) and mainly consists of thermal plasma with a temperature of $\SI{e7}{\kelvin}$ to $\SI{e8}{\kelvin}$ emitting bremsstrahlung at X-ray wavelengths.
In addition, the ICM contains a non-thermal component consisting of cosmic rays and magnetic fields. This non-thermal component can be observed with radio telescopes as extended radio synchrotron emission \citep[for reviews see ][]{2012A&ARv..20...54F,diffuse_radio}.

Megaparsec-scale diffuse radio emission in clusters is generally classified into two categories: halos and relics. 
Radio relics typically have an elongated morphology with a relatively high polarisation fraction \citep[up to $70\%$ at GHz frequencies, e.g.][]{2022A&A...659A.146D} and are generally located in the cluster periphery \citep[][and references therein]{diffuse_radio}.
Radio halos are diffuse extended sources, with the radio emission approximately following the thermal X-ray emission from the ICM. They are ubiquitous at low frequencies in massive merging clusters and have a typical extent of about 1\,Mpc.
The origin of the radio-emitting cosmic ray electrons producing radio halos might trace complex re-acceleration mechanisms in the ICM \citep[e.g. ][]{brunettijones2014}.
For radio halos, there are two proposed physical mechanisms to explain how electrons can be (re-)accelerated to relativistic energies.

The first, the `hadronic model', predicts that cosmic ray electrons can be produced from collisions between cosmic ray protons and thermal protons \citep{blasi_cola_crp,brunettiblasi2005,pinzke2017,2024MNRAS.527.1194K}.
However, this explanation is disfavoured as being the primary mechanism due to the non-detection of gamma-ray emission \citep{2017MNRAS.472.1506B,gamma_nondetection,2024A&A...688A.175O} and the existence of radio halos with a very steep spectrum \citep{2008Natur.455..944B}.

A second model, the `turbulent re-acceleration' model, describes the formation of these halos as being a consequence of mergers of galaxy clusters \citep[e.g.][]{2001MNRAS.320..365B,2001ApJ...557..560P,2005MNRAS.357.1313C,2006MNRAS.369.1577C,2007MNRAS.378..245B,cwperturbed}. 
Galaxy cluster mergers inject turbulence into the ICM, which causes an underlying population of relativistic seed electrons to be re-accelerated to higher relativistic energies \citep{2001MNRAS.320..365B,2001ApJ...557..560P}. 
This re-accelerated plasma emits synchrotron emission, which contains a high-energy cut-off, above which synchrotron and inverse Compton losses dominate over the rate of re-acceleration \citep{1999ApJ...520..529S,lowfreq_radX,radio_halos_cassano}.
More energetic mergers dissipate larger amounts of kinetic energy through turbulence, which causes re-acceleration to dominate over radiative losses up to a higher frequency \citep{2006MNRAS.369.1577C,2023A&A...680A..30C}. 
As the merger energetics depend on the mass of the colliding clusters, massive systems are able to re-accelerate electrons to higher energies, producing radio halos with a spectral break at higher frequencies.
At a fixed frequency, this generally causes higher-mass clusters to host radio halos with significantly flatter spectra and lower-mass clusters to host radio halos with steeper spectra, although in some cases high-mass clusters can still host steep spectra, possibly due to a high impact parameter or mass ratio during the merger \citep[e.g. Abell 521, ][]{2024ApJ...962...40S}.
In particular, such re-acceleration models predict that a significant amount of galaxy clusters host halos with an ultra steep spectrum ( $\alpha < -1.5$), which are restrictively difficult to detect at radio frequencies $> 1$ GHz \citep{2008Natur.455..944B,radio_halos_cassano,dallacasa_1p4GHz,macario_a697}.
However, with the advent of low-frequency radio observatories (e.g. LOFAR, MWA), observers have been able to detect this population of ultra steep spectrum radio halos \citep[e.g.][]{rvw_2012,venturi2017,wilber2018,2021A&A...654A.166D,bruno2021,2021PASA...38...31D,2022A&A...666A...3E,2024A&A...689A.218P}. %
The LOw Frequency ARray \citep[LOFAR;][]{lofar} in particular is  well suited to detecting radio halos associated with clusters of relatively lower mass ($\lesssim 5 \times 10^{14} M_\odot$), which have previously gone unnoticed when observing at higher radio frequencies.

The common property that large-scale diffuse radio emission holds is that while models can describe the acceleration of particles to high-energy cosmic rays, they require an underlying `seed' population of electrons to be re-accelerated.
Radio galaxies in the cluster environment are believed to play an important role in the origin of this seed electron population \citep[e.g.][]{2001MNRAS.320..365B,2003ApJ...584..190F,2017NatAs...1E...5V,stuardi2019,2024Galax..12...19V}.
Other potential options include deposition through a hadronic process \citep{2011MNRAS.410..127B} or previous structure formation shocks \citep{2013MNRAS.435.1061P}. 

Observations have also revealed smaller-size (a few hundred kiloparsecs) diffuse radio sources in galaxy clusters with ultra steep spectra, and often with a complex morphology. A proposed scenario is that these sources are formed by fossil plasma from ancient Active Galactic Nucleus (AGN) outbursts that have since aged and subsequently been re-energised \citep[e.g.][]{slee2001,fossil_plasma,diffuse_radio,2020A&A...634A...4M}. This can be caused by adiabatic compression due to large-scale structure formation shocks \citep[radio phoenixes;][]{fossil_plasma} or via more complex plasma interactions \citep[such as Gently Re-energised Tails;][]{greet}. Radio phoenixes have an irregular filamentary structure, as shown by simulations \citep{phoenix_simulation,fossil_plasma} and observations \citep[e.g.][]{2015dgasp,a2256phoenix,2023arXiv230914244R}. This plasma often has a relatively low surface brightness at gigahertz radio frequencies and a particularly steep spectrum above a certain frequency. When an optical counterpart can be identified, the spectral index at the location of the counterpart typically flattens \citep{2020A&A...634A...4M}, which suggests that there is a connection between the AGN and the radio phoenix.
U
\begin{figure*}
    \centering
    \includegraphics[width=0.75\textwidth]{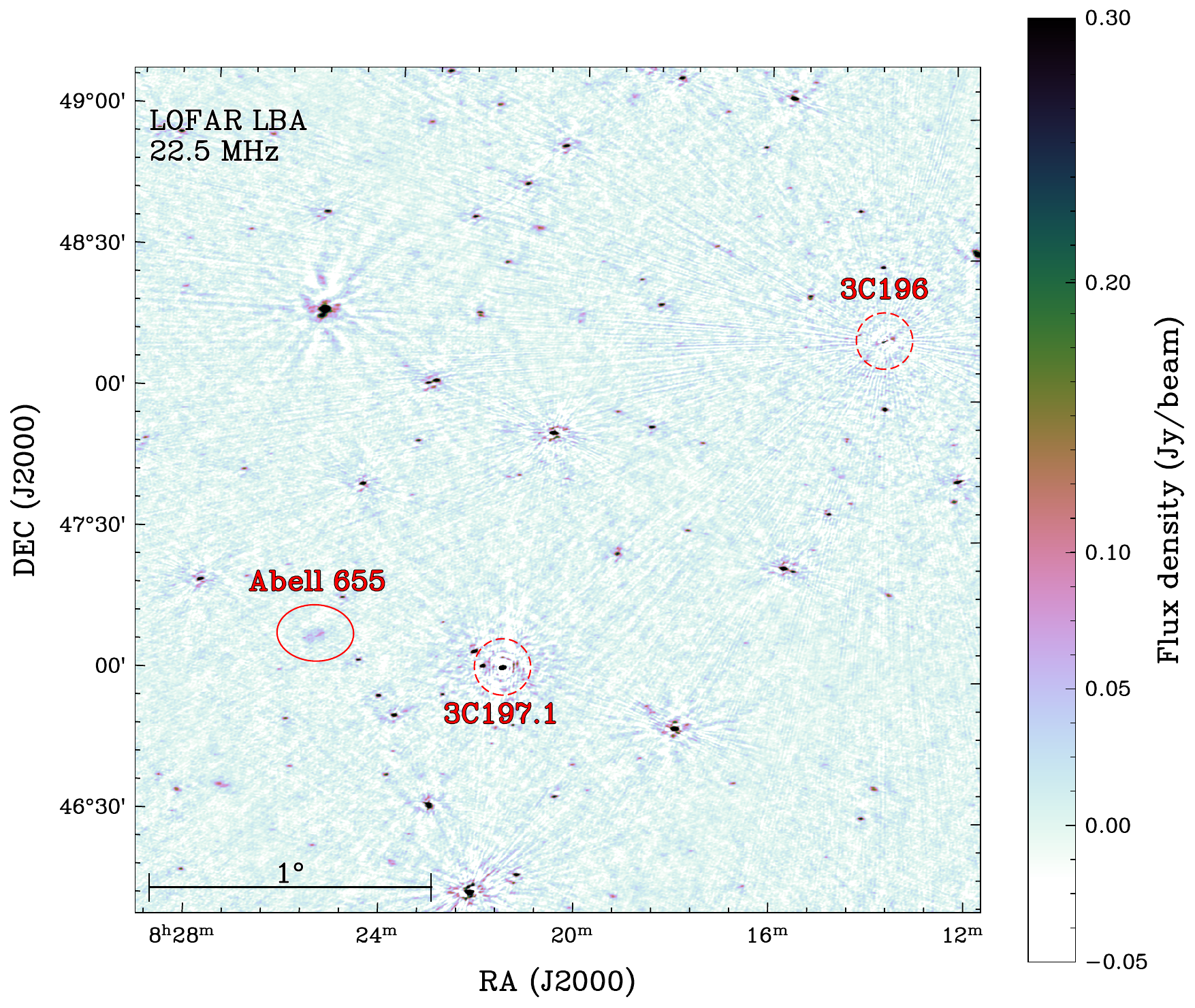}
    \caption{Widefield image at 15--30\,MHz of the area around 3C\,196, The source 3C\,196 itself has been subtracted from the data, which would otherwise significantly lower the dynamic range of the image. }
    \label{fig:finder}
\end{figure*}

The past decade has seen enhanced interest in the non-thermal component of galaxy clusters, driven by discoveries of diffuse emission on large scales, such as radio bridges \citep{govoni_sci,botteon1758} and pervasive emission in cluster outskirts \citep[e.g.][]{shweta2020,megahalos_nature,rajpurohit2021,2022SciA....8.7623B,bruno2023}, exploiting observations at lower frequencies than was possible before.
These gigantic radio structures may trace turbulence and shocks in the outskirts of galaxy clusters \citep{2020PhRvL.124e1101B,2024ApJ...961...15N}.
However, observations of galaxy clusters below 30\,MHz are extremely scarce \citep{rvw_2012,2024A&A...688A.175O} due to the severe perturbing effects of the ionosphere \citep[e.g.][]{2018A&A...615A.179D}. This limits our understanding of the spectral shape of the non-thermal plasma in galaxy clusters at such low frequencies.

In this work, we study the diffuse radio emission permeating the low-mass galaxy cluster Abell\,655 located at $z=0.129$, whose detection was previously reported with LOFAR by \cite{planck_lofar}. In \autoref{tab:abell_data}, we list some of the properties of the cluster.
Extended radio emission of this cluster was serendipitously detected in an ultra low frequency observation carried out with LOFAR of the field containing the flux density calibrator 3C\,196, which is indicative of the emission having a very steep radio spectrum, as can be seen in Fig. \ref{fig:finder}. In this work, we characterise the emission present in this cluster and investigate the connection between large-scale diffuse radio emission and old AGN outbursts.
This paper uses a flat $\Lambda$-CDM cosmology from \cite{planck18}. Spectral indices are defined by $S_\nu\propto \nu^\alpha$.

\begin{table}[!h]
    \centering
    \caption{Properties of Abell 655.}
    \label{tab:abell_data}
    \begin{tabular}{ll}
    \centering Abell 655 & \\
    \hline \hline & \\
    Planck SZ & PSZ G172.63+35.15 \\
    RA\textsubscript{J2000}$\,^{(1)}$ &  08:25:30.3 \\
    DEC\textsubscript{J2000}$\,^{(1)}$ & +47:07:48 \\
    z\textsubscript{spec}$\,^{(1)}$ & 0.129 \\
    $L$\textsubscript{0.7-2 keV}$\,^{(1)}$ & \SI{3.83\pm0.55e44}{\erg\per\second} \\
    $R_{500}\,^{(2)}$ & \SI{1.1\pm 0.03}{\mega\parsec} \\
    $M_{500}\,^{(2)}$ & \num{3.9\pm 0.32e14} $M_\odot$ \\
    $c$\textsubscript{Chandra}\,$^{(3)}$ & \num{0.184 \pm 0.009} \\
    $w$\textsubscript{Chandra}\,$^{(3)}$ & \num{0.0201\pm 0.0021} \\
    \hline &
    \end{tabular}
    \tablefoot{$^{(1)}$Data obtained from \cite{xray_a655}; $^{(2)}$data obtained from \citep{planck2016}; $^{(3)}$data obtained from \citep{cluster_states}. The $c$,$w$ parameters are described in \citep{cwperturbed}}
\end{table}

\section{Methods}
\label{sec:methods}

\begin{table}
    \centering
    \caption{Properties of the radio images presented in Fig. \ref{fig:sixpanels}.}
    \begin{tabular}{c|c|c|c}
         & Frequency & Noise & Resolution  $^{(1)}$ \\ \hline
        LBA Low & 15--30~MHz & 17 mJy\,beam$^{-1}$ & 40''$\times$ 24'' \\
        LBA High & 30--60~MHz & 3.2 mJy\,beam$^{-1}$ & 19''$\times$ 12'' \\
        HBA & 120--167~MHz & 0.28 mJy\,beam$^{-1}$ & 24''$\times$22'' $^{(2)}$\\
        GMRT & 550-750~MHz & 9.5 \si{\micro\jansky}\,beam$^{-1}$ & 8''$\times$3.4''
    \end{tabular}
    \tablefoot{$^{(1)}$ The resolution reported here represents the major axis $\times$ minor axis of the restoring beam. $^{(2)}$ The resolution is set by a $uv$-taper that matches a physical size of 25 kpc at the cluster redshift }
    \label{tab: properties}
\end{table}

\subsection{LOFAR data}

For this work, we used an 8\,hr LOFAR Low Band Antenna (LBA) observation between 12--60\,MHz centred on the bright radio source 3C\,196, which is located $\SI{2.2}{\degree}$ from Abell\,655.
The size of the primary beam is 5.8 degrees at 23~MHz, which means that the source is well within the primary beam.
The observation was taken on October 28, 2020, between midnight and 08:00 UTC.
The data processing started with the removal of the signal from the bright radio sources Cygnus~A and Cassiopeia~A using the technique described in \cite{lofar_selfcal}.
Next, we applied solutions for instrumental effects, in particular the bandpass response and polarisation alignment (the phase offset between the X- and Y-feeds of each antenna for an unpolarised signal) of the individual stations.
Next, we applied the bandpass response and polarisation alignment correction to our data \citep{2019A&A...622A...5D}.
As these effects are largely time independent, we could use solutions that were obtained during an earlier calibration run of 3C\,196 for this step.
After this step, we continued using the \texttt{facetselfcal} script described in \cite{facetselfcal}, where we used a perturbative calibration strategy to solve for Faraday rotation and antenna-based gains (both for phases and amplitudes).
For the most distant antennas (RS310 and RS210), we detected a differential rotation measure (with respect to the LOFAR core) of at most $\lesssim$ 0.01 rad m$^{-2}$, which is considered relatively low \citep{2019A&A...622A...5D}.
For this direction-independent calibration step, we used a high-resolution reference model from \cite{highres_lba}, with the fluxes fixed to the \cite{scaife_heald} flux density scale.\footnote{Even though the \citep{scaife_heald} flux density scale is formally only valid down to 30~MHz, the spectrum of 3C\,196 seems relatively well constrained down to 10\,MHz (see Fig.~2 of \citeauthor{scaife_heald}).}
After calibration, %
the 3C196 model was subtracted from the visibilities, and the full field was imaged with WSClean \citep{wsclean,msmfs}.

\begin{figure}
    \centering
    \includegraphics[width=0.9\linewidth]{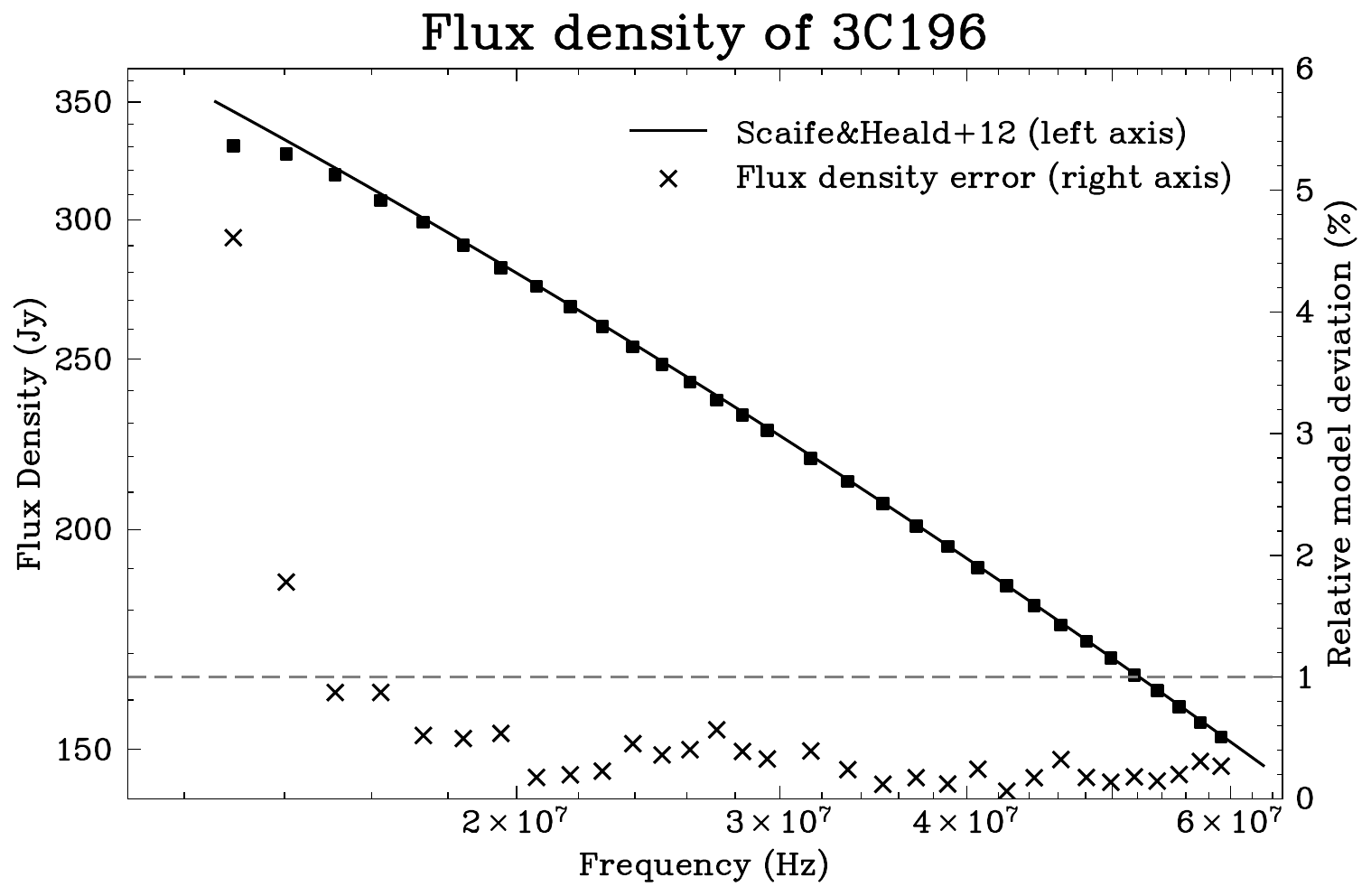}
    \caption{Flux density of 3C\,196, as measured after initial calibration. As the data was calibrated against a \citep{scaife_heald} flux scale model of 3C\,196, the deviation from the model should be minimal. Deviations from the flux scale indicate instability in the calibration, which means that the corresponding data should be discarded. This figure shows that data below 15\,MHz is too unreliable to be used.}
    \label{fig:3C196_SH}
\end{figure}

To verify the calibration quality, we used the dataset before the subtraction of 3C196. We compared the flux density of 3C\,196 measured from the in-band spectral images (see Fig. \ref{fig:3C196_SH}) with the flux density as provided by \cite{scaife_heald}, which was imposed on the data during calibration. Significant deviations from \cite{scaife_heald} imply that our calibration was unable to properly correct the data for ionospheric effects. This can happen when the phase corruptions vary on time and frequency intervals shorter than the calibration solution intervals can handle, and this can occur because we are observing close to the plasma cut-off frequency.
Figure \ref{fig:3C196_SH} reveals that the measured flux densities below 15~MHz fall more than 1\% below the model flux densities, increasing up to ${\sim}5\%$ at 12\,MHz. Given that 3C\,196 is by far the brightest source in the field of view and noting that any subsequent direction-dependent calibration steps will have to work in an even lower signal-to-noise regime, we decided to discard the data below 15\,MHz from this point on.

\begin{figure*}
    \centering
    \includegraphics[width=0.45\linewidth]{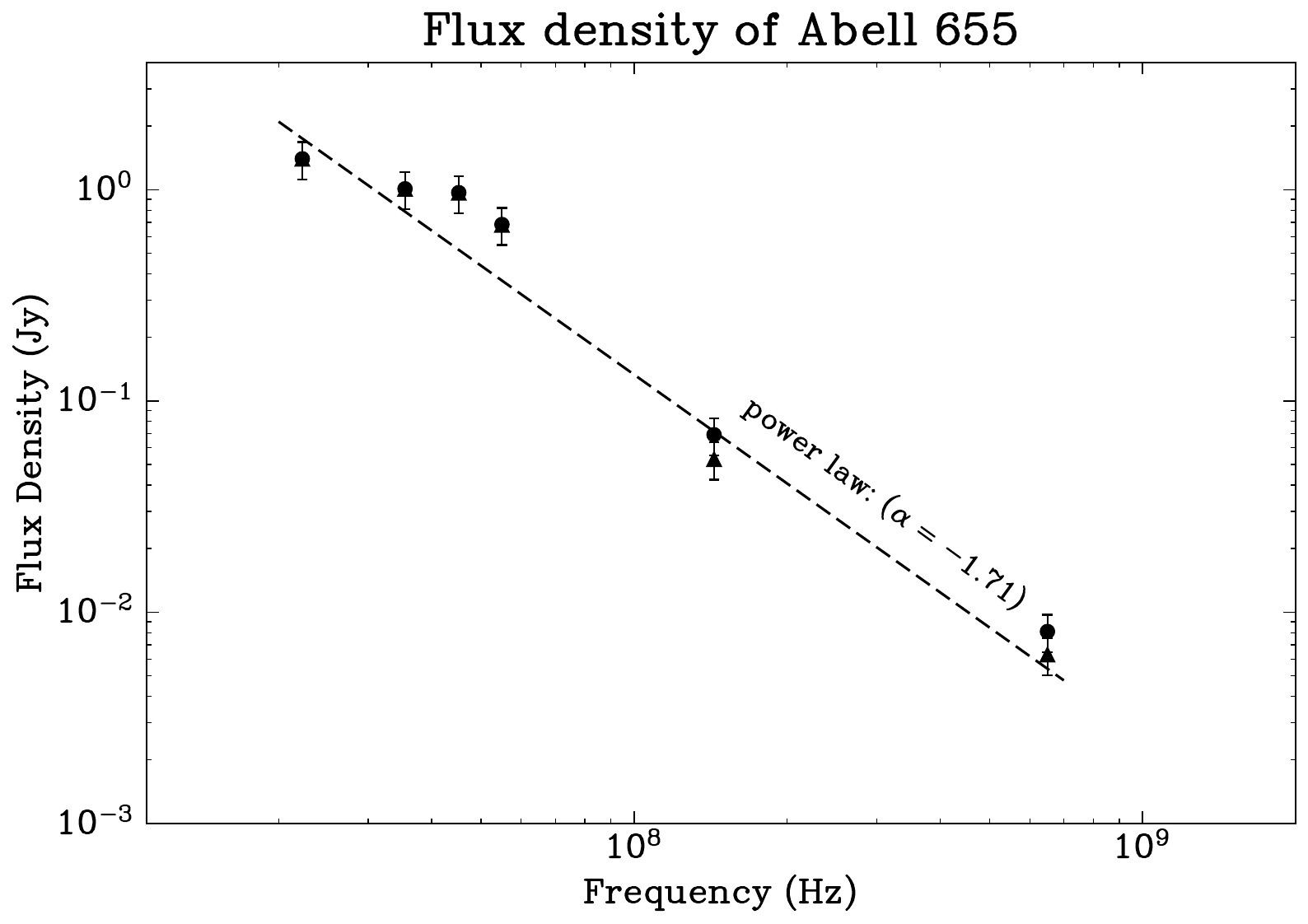}
    \includegraphics[width=0.45\linewidth]{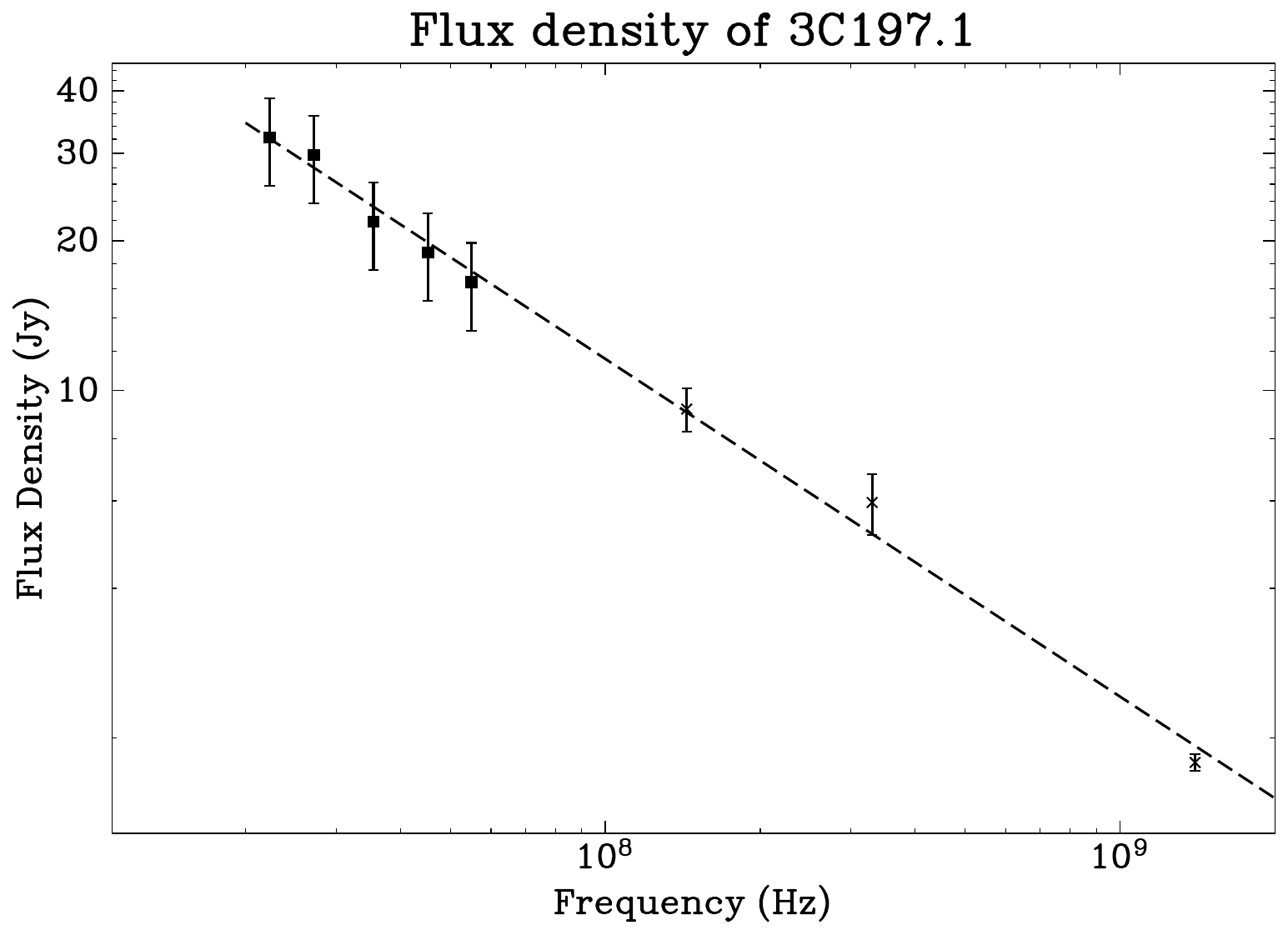}
    \caption{Integrated spectra of the cluster and 3C\,197.1. 
    Both flux densities with (circles) and without (triangles) point source subtraction are reported in the case of Abell\,655.
    We note that for the LBA flux densities, the contribution of point sources to the total flux density is negligible,%
    which means that the data with and without subtraction of point sources are indistinguishable within the uncertainties.
    The dotted line represents a power-law spectral profile fitted to the point-source subtracted flux density, and it gives an indication of the steepness of the spectrum of A655, with a spectral index of $\alpha=-1.71\pm0.16$. 
    The radio spectrum of 3C\,197.1 is that of a regular AGN, with a spectral index of $\alpha = -0.68 \pm 0,02$, which allows us to determine the validity of our LOFAR flux density measurements. %
    }
    \label{fig:int_spec}
\end{figure*}
The next step consisted of direction-dependent calibration.
At low frequencies, the corrections for the ionosphere derived during the direction-independent calibration are only valid in a relatively small region of sky near a calibrator source.
The size of this region depends on the baseline length and ionospheric conditions; for core stations, this region can be as big as the field of view, while for the most distant remote stations, this could be as small as several tens of arcminutes.
Therefore, the field was divided further into seven separate facets, and each facet was calibrated individually following an approach similar to \cite{tmpDeca} (hereafter G24).
Each facet was separately split off from the rest of the field (peeling), and after this, each facet was calibrated for differential Faraday rotation and differential phases.
The data was re-imaged with the corrections applied using \texttt{WSClean}. 

Finally, an image of Abell\,655 was made with WSClean. In order to do this, the sources outside of a $1.7\times1.7$ degree$\,^{2}$ box around Abell 655 were subtracted from the visibilities before self-calibration was performed.
The flux density scale was verified with a nearby source, 3C\,197.1, as can be seen in the right panel of Fig. \ref{fig:int_spec}. 
A plot of the spectrum of the nearby source 3C197.1 (indicated in Fig. \ref{fig:finder}) is shown along with data points from LoTSS DR2 \citep[144~MHz;][]{lotss2}, WENSS \citep[325~MHz,][]{wenss}, and NVSS \citep[1.4~GHz,][]{nvss}. 
The integrated spectrum looks similar to that of a normal AGN, with a spectral index of $\alpha = -0.68\pm 0.02$, which suggests that the flux density scale is reliable down to 15~MHz.

In this work, we also employ 120--168\,MHz LOFAR High Band Antenna (HBA) data that is available via the LoTSS survey  \citep[Data release 2,][]{lotss2} and was published earlier in \cite{planck_lofar}. We used the reprocessed images of the cluster presented in the latter work, re-imaged with a 11\arcsec{} $uv$-taper, corresponding to 25\,kpc in the cluster rest frame.

\subsection{Giant Metrewave Radio Telescope data reduction}
\begin{figure*}[ht]
    \centering
    \includegraphics[width=\linewidth]{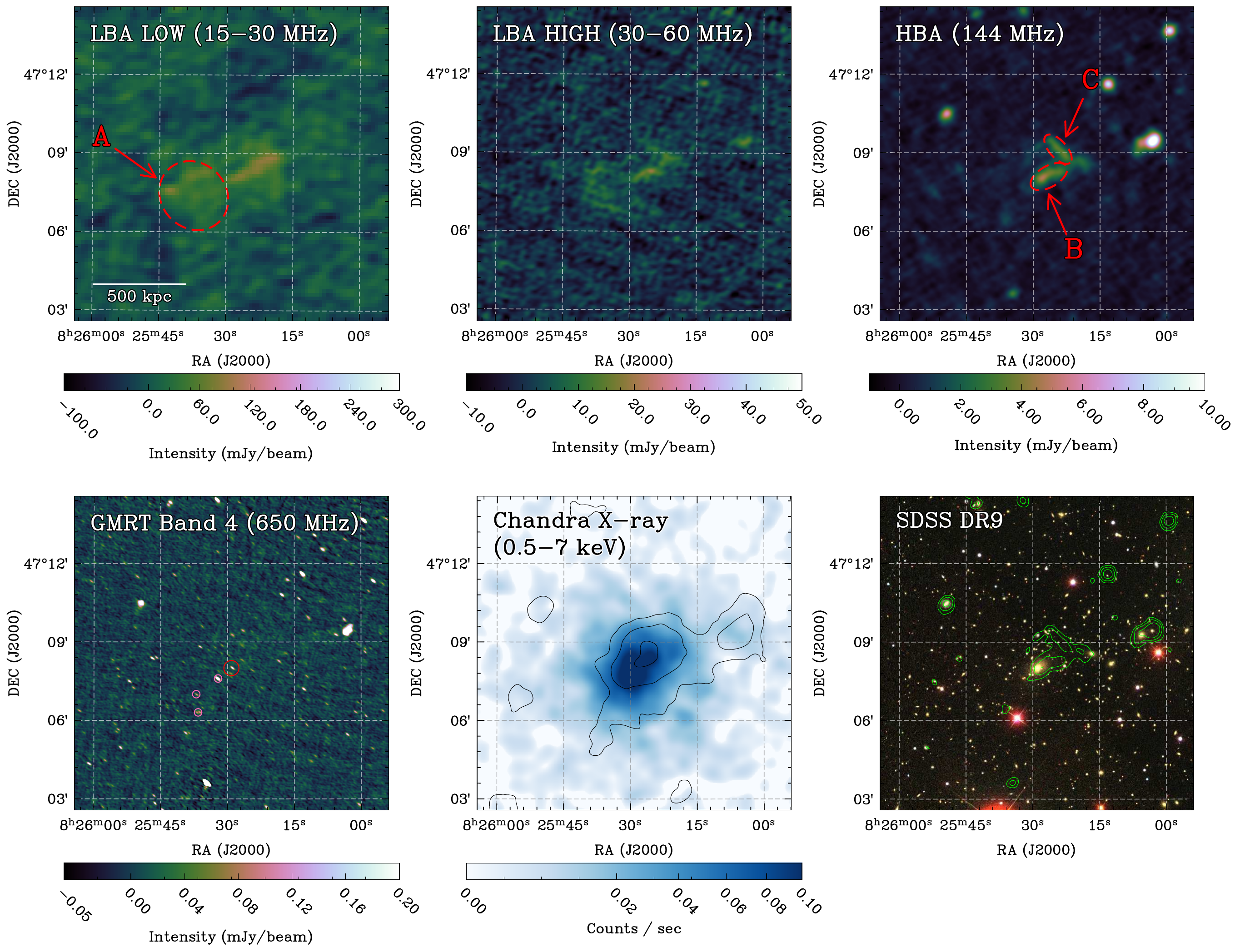}
    \caption{Images of Abell\,655. \textit{Top left: } LOFAR LBA (15-30 MHz) image.
    \textit{Top centre: } LOFAR LBA (30-60 MHz). \textit{Top right: } LOFAR HBA (120--167 MHz), tapered to 25 kpc (rest frame of the cluster). \textit{Bottom left: } GMRT band 4 (450--850 MHz), tapered to 12 arcseconds. The BCG of Abell\,655 is marked with a red circle. The small pink circles indicate AGN that were subtracted from the integrated flux density of region A. \textit{Bottom centre: } \textit{Chandra} X-Ray \citep[ObsID 15159; 7.97 ks integration time; 0.5--7 keV,][]{planck_lofar} with GMRT 650\,MHz overlay (compact sources subtracted). Contours are at $[1,2,4,8]\times \sigma_{RMS}$, with $\sigma_{RMS}=\SI{0.2}{\milli\jansky\per\beam}$. \textit{Bottom right:} SDSS DR9 colour image, with LOFAR HBA contours. Contours are at $[4,8,16] \times \sigma$, with $\sigma=\SI{0.3}{\milli\jansky\per\beam}$. Regions corresponding to the regions used in Fig. \ref{fig:spix}are marked with dashed red lines in the LOFAR images. }
    \label{fig:sixpanels}
\end{figure*}

In addition to the LOFAR data, 5.5 hours of upgraded Giant Metrewave Radio Telescope (uGMRT) band 4 (centred at 650~MHz) data were obtained, with a total bandwidth of 200~MHz (proposal ID 43\_023, PI R. Cassano). The data were recorded with 2048 channels, with a channel width of 97.7~kHz per channel and an integration time of 5.4 seconds.
The data were bookended by calibrator scans of 3C48 and 3C147.
The data reduction was performed using the SPAM pipeline \citep{2009A&A...501.1185I,spam}, which performs both direction-independent and direction-dependent calibration on the GMRT data.
For this, we divided the full 200~MHz bandwidth into four chunks of 50~MHz bandwidth, which were all separately calibrated.
After running the SPAM pipeline, the four chunks were imaged together at full (8.09 arcsec $\times$3.47 arcsec) resolution using \texttt{WSClean}, with a Briggs weighting of 0. The resulting image is presented in the bottom-left of Fig. \ref{fig:sixpanels}, and it has an RMS noise of \SI{11}{\micro\jansky\per\beam}.
Subsequently, another image was made with an inner $uv$-cut of 3000 $\lambda$ and with a Briggs weighting of -0.5. We used this image to get a high-resolution model with exclusively compact sources and without any diffuse emission.
The compact sources were then subtracted from the visibilities by predicting the high-resolution model and subtracting the result from the original visibilities.
The data were re-imaged with a Gaussian $uv$-taper of 20\arcsec and with a Briggs weighting of 1, allowing for better detection of diffuse emission.
This image is presented as a contour map in the lower centre image in Fig. \ref{fig:sixpanels}.

\section{Results}
In Fig. \ref{fig:sixpanels} we present images of Abell\,655, and in Fig. \ref{fig:int_spec}, we present the integrated spectrum of Abell\,655 (left).
The image properties of the images presented in Fig. \ref{fig:sixpanels} are shown in \autoref{tab: properties}.

At frequencies between 15--30\,MHz (`LBA LOW'), we detected extended emission in the cluster region with a total extent of around 5\arcmin{} (\SI{700}{\kilo\parsec}). The diffuse emission in the LBA LOW image seems to consist of two parts: a diffuse region of emission to the east, marked with `A', and an elongated brighter structure west of this region.
At higher frequencies (30--60\,MHz, `LBA High'), we generally detected the same structures as in the LBA LOW image. The radio appearance of the cluster, however, changes in the 120--168\,MHz HBA image. There is barely any emission visible from region A at 25 kpc resolution. The emission in the western part of the cluster consists of two elongated structures labelled `B' and `C'. These structures can also be recognised in the LBA HIGH image, while in the LBA LOW image, B and C appear partly blended.

When comparing the HBA image to the optical overlay of SDSS DR9 \citep{sdss3}, we observed that the emission corresponding to region B seems to originate at the brightest cluster galaxy (BCG).
In the GMRT Band 4 image, we detected compact radio emission from the nucleus of this galaxy (marked with a red circle), indicating that region B is likely associated with emission from a radio AGN with a `tailed' morphology.
The origin of the emission corresponding to region C is less clear, as there is no obvious optical counterpart, although a potential radio counterpart is visible when comparing the full resolution GMRT 650~MHz map with the SDSS overlay (see Fig. \ref{fig:gmrt_sdss}). If this compact AGN visible in the 650~MHz map is indeed related to the elongated structure C, the morphology resembles that of a tailed radio source.

In the integrated spectrum of Abell\,655, both the total spectrum and point-source subtracted spectrum are reported.
The point sources were subtracted from the LBA images by measuring the flux density of the three AGN detected within region A in the HBA images and scaling the measured flux density to the relevant LBA frequencies with a canonical spectral index of $\alpha=-0.7$.
The three AGN are marked in Fig. \ref{fig:sixpanels} with pink circles in the GMRT image, as that is where the AGN are most clearly visible.
The integrated radio spectrum of Abell\,655 shows clear signs of spectral flattening at frequencies below 144~MHz (flattening to $\alpha=-0.73\pm 0.06$), while the spectrum is extremely steep ($\alpha \approx -2.1$) above 144~MHz.
\begin{table*}[]
    \centering
    \caption{Flux densities in the three regions (labels match with the regions in Fig. \ref{fig:sixpanels}).}
    \begin{tabular}{l|c|c|c}
         & A & B & C \\ \hline
        15--30~MHz & $(5.3\pm0.17)\times\num{e-1}$  & $(1.8\pm0.6)\times\num{e-1}$ & $(1.1\pm0.4)\times\num{e-1}$ \\
        30--60~MHz & $(2.4\pm0.9)\times\num{e-1}$ & $(1.0\pm0.4)\times\num{e-1}$  & $(6.2\pm2.8)\times\num{e-2}$\\
        120--167~MHz & $(1.1\pm0.2)\times\num{e-2}$ & $(1.4\pm0.3)\times\num{e-2}$ & $(9.5\pm1.9)\times\num{e-3}$ \\
        550--750~MHz & --- & $(1.1\pm0.2)\times\num{e-3}$ & $(5.5\pm1.1)\times\num{e-4}$
    \end{tabular}
    \tablefoot{The flux densities are reported in Jansky. Region A has no reliable emission in the GMRT (550-750~MHz) frequency range and is therefore omitted from the table. }
    \label{tab:fluxdensities}
\end{table*}

\begin{figure*}
    \centering
    \includegraphics[width=0.48\linewidth]{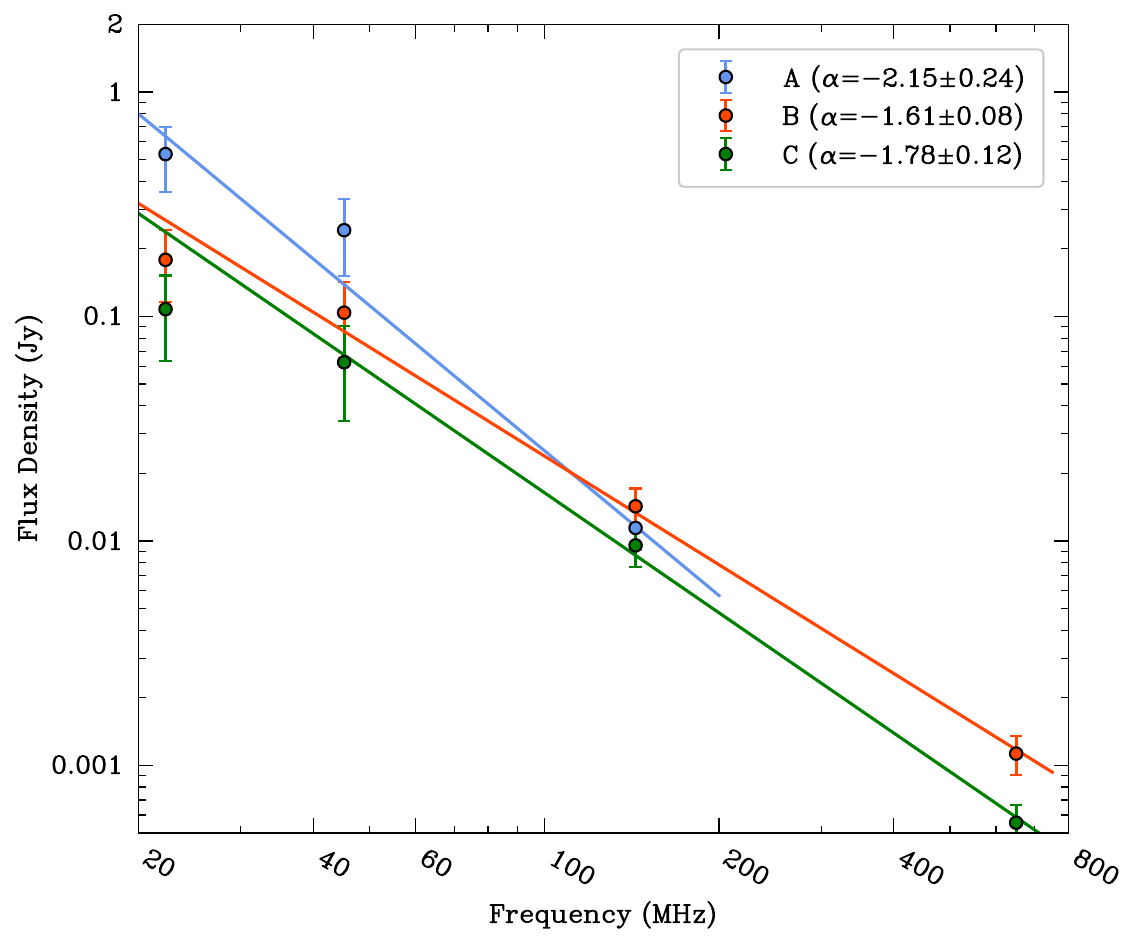}
    \includegraphics[width=0.48\linewidth]{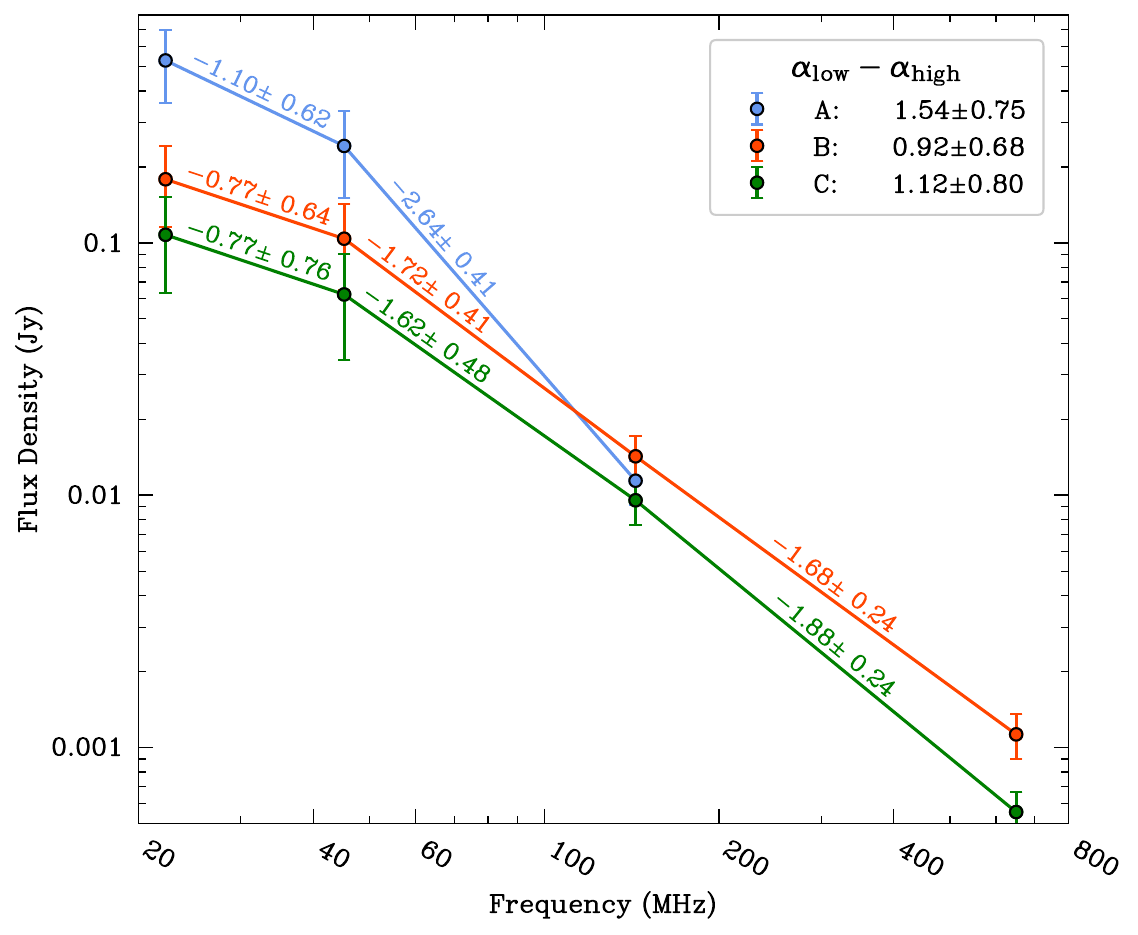}
    \caption{Spectral energy distribution of three regions, showing that only region A has signficant ($> 2\sigma$) curvature. Labels in the plot correspond to the regions labelled in Fig. \ref{fig:sixpanels}. \textit{Left: } Lines are a least-squares polynomial fit through the data in order to give an indication of the steepness of the spectra. The spectra indicate that region A is the steepest. \textit{Right:} Piecewise spectral indices of the three regions, including error. In the top-right corner, the difference between the spectral index at the highest frequency and the spectral index at the lowest frequency is given.  }
    \label{fig:spix}
\end{figure*}

Since region A is only clearly detected in the LBA images, this indicates it has a steeper spectrum than the components B and C. 
To determine the radio spectra of these components, we extracted the flux densities in the regions indicated in Fig. \ref{fig:sixpanels} and reported them in \autoref{tab:fluxdensities}. 
The results are shown in Fig. \ref{fig:spix}. 
The radio spectral index is taken by fitting a single power law through the four flux measurements (weighted by the inverse of the squared standard error).
Region A had no significant diffuse emission in the GMRT map, and therefore the 650~MHz data point is omitted for region A.
The results confirm that the region marked with `A' contains ultra steep spectrum emission, with a spectral index of $\alpha=-2.15\pm0.24$, which is significantly steeper than the emission from both region B ($\alpha = -1.61 \pm 0.08$) and region C ($\alpha=-1.78 \pm 0.12$). 
From Fig. \ref{fig:spix}, it can be seen that the spectra of regions B and C have a hint of curvature, whereas for region A, the relatively poor power-law fit indicates the presence of a potentially significant curvature.
This can be verified by comparing the spectral index between the highest two frequencies (144~MHz and 45~MHz for region A, 144~MHz and 650~MHz for the other two regions)  and the spectral index between the lowest two frequencies (45~MHz and 23~MHz).
These spectral indices are reported in Fig. \ref{fig:spix} (right panel).
Using the difference between the spectral index of the highest two and lowest two frequencies, we can conclude that only region A hosts a significant curvature.

In addition to the radio images, we present a radio overlay on a \textit{Chandra} X-ray (0.5--7~keV, obsID \texttt{15159}, 7.97 ks) image (Fig.~\ref{fig:sixpanels}, bottom-middle panel).
The \textit{Chandra} observation was reduced using the procedure described in \citep{planck_lofar} with the \texttt{CIAO} software package \citep{2006SPIE.6270E..1VF}. The \textit{Chandra} data is shown with a contour overlay of the source-subtracted GMRT image, which reveals diffuse radio emission coinciding with the X-ray emission, which is a typical sign of a radio halo \citep[e.g.][]{diffuse_radio}.
The radio emission appears to be slightly offset in the direction of the two tailed structures (B and C).
In the source-subtracted and tapered images presented in \cite{planck_lofar}, a similar region of diffuse emission can be seen in the galaxy cluster.

\begin{figure}[h!]
    \centering
    \includegraphics[width=0.8\linewidth]{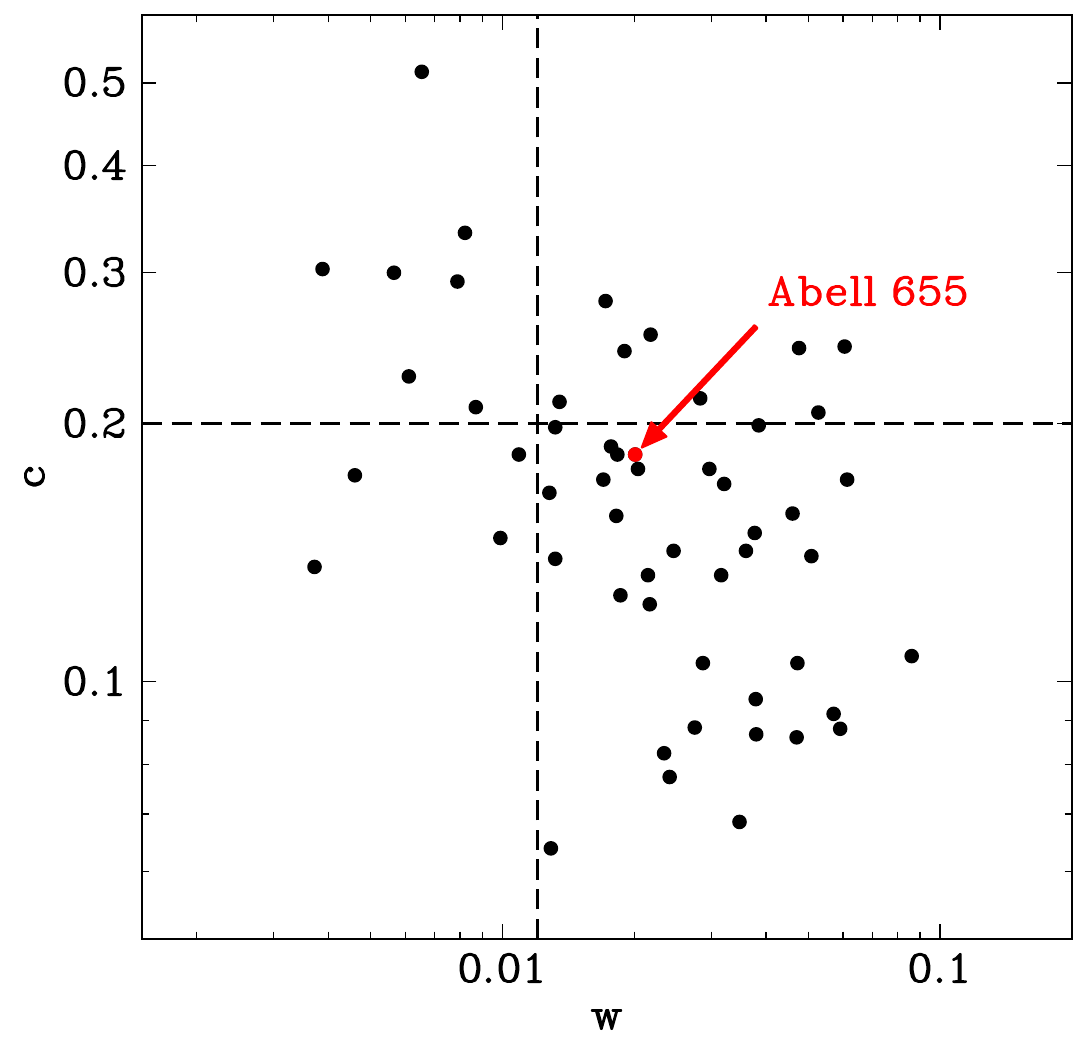}
    \caption{Comparison of the concentration parameter ($c$) and the centroid shift ($w$) of the galaxy clusters with radio halos in the Planck-LOFAR catalogue \citep{planck_lofar}. The dashed lines separating disturbed (bottom-right quadrant) from non-disturbed (top-left quadrant) galaxy clusters correspond to $w=0.012$ and $c=0.2$ \citep{cwperturbed}.  }
    \label{fig:cw}
\end{figure}

\begin{figure}
    \centering
    \includegraphics[width=0.9\linewidth]{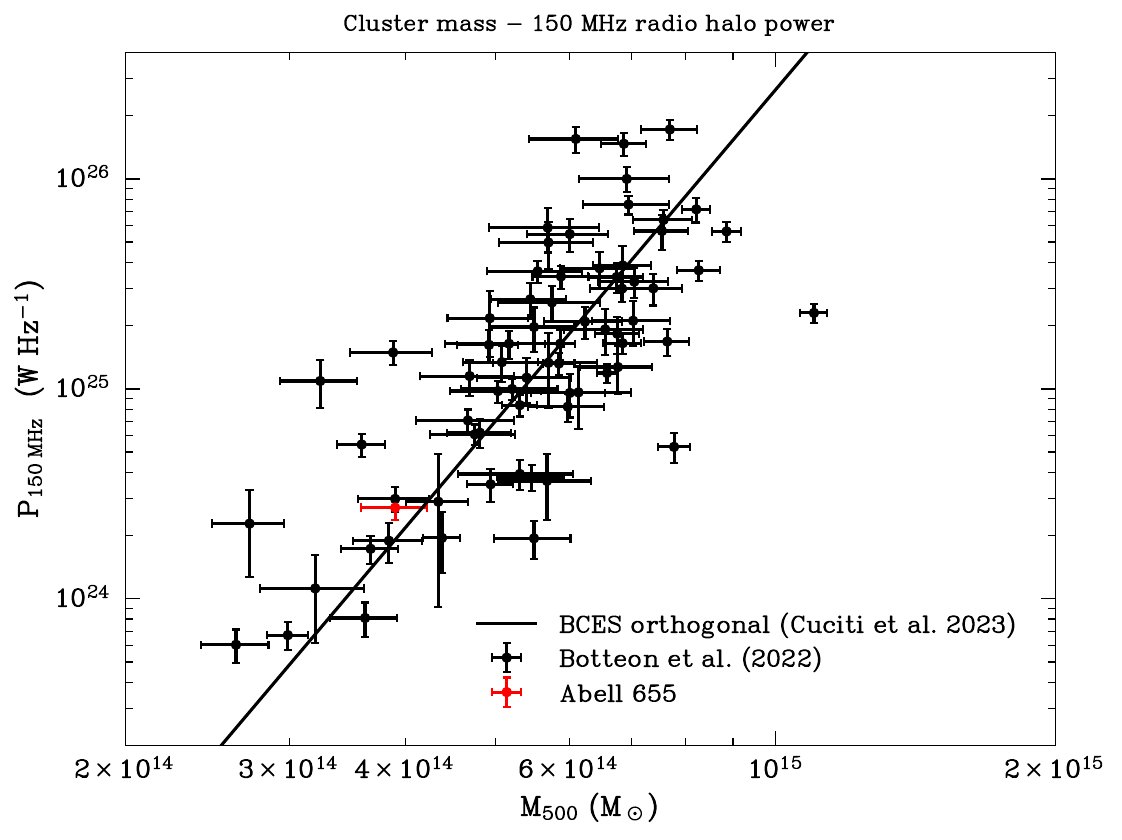}
    \caption{Power-mass plot of radio halos. Black points are taken from \cite{planck_lofar}, and the solid line represents the BCES fit from \cite{2023A&A...672A..43C}. The red point is Abell 655, and the halo power was obtained from \citep{planck_lofar}.}
    \label{fig:halopower}
\end{figure}

\section{Discussion}
\subsection{Origin of the radio emission}

The radio observations presented in this work report diffuse radio emission with a physical extent of about 700\,kpc in the cluster Abell\,655. %
In general, there are several possibilities for classifying and determining the origin of extended steep-spectrum emission in clusters. 
This type of radio emission can be associated with a radio halo resulting from inefficient re-acceleration by turbulence, originate from ageing radio plasma in AGN lobes and tails, or manifest as re-energised radio plasma through mechanisms such as adiabatic compression (i.e. radio phoenixes). The integrated spectrum of the cluster emission (see Fig. \ref{fig:int_spec}) is relatively flat at very low frequencies but becomes extremely steep at higher frequencies. This could indicate energy losses or the presence of multiple separate radio components, each with its own individual spectral shape.

The GMRT image shows that the radio emission at 650\,MHz has a roughly spherical morphology, similar to the X-ray emission. This is expected for a radio halo \citep{xray_radio_halo,xray_radio_halo2}.
This emission was also classified as a halo  by \cite{planck_lofar} at 144~MHz, but we are unable to confidently distinguish the halo from the other radio sources in the LOFAR LBA image.
The halo power of Abell 655 is slightly above the best fit correlation, based on its mass (Fig. \ref{fig:halopower}). 
The main reason for this is the limited sensitivity and resolution of LBA observations, which severely complicates source subtraction.
Without source subtraction, the other components (most notably the two tails, as they are coincident with the radio halo) are dominant over the radio halo.

Figure \ref{fig:sixpanels} shows that the origin of the emission in Abell\,655 is unlikely to be from one single source -- instead it seems that the emission labelled as `A' brightens significantly in the decametre band, whereas the sources labelled as `B' and `C' appear to host a relatively flatter spectrum, as can also be seen in Fig. \ref{fig:spix}. 
The source marked with `A' has no clear optical counterpart. 
Given the steep ($\alpha=-2.15\pm0.24$) and curved radio spectrum of A, we can provide two potential origins for this emission:
Firstly, the source could be emission tracing old radio lobes, which are possibly re-energised.
These old radio lobes trace old outbursts of AGN, inflating bubbles of AGN plasma that buoyantly move to the cluster periphery \citep{2021NatAs...5.1261B}.
If this is indeed the case, we expect that deeper observations will reveal the emission to have a complex morphology, such as those observed for other radio phoenixes \citep[e.g.][]{2015dgasp,a2256phoenix}.
Alternatively, this emission could also be part of the radio halo, making the radio halo an ultra steep spectrum radio halo.
It is, however, difficult to make a confident characterisation of the nature of the diffuse emission, as the fidelity of the LBA image of region A is limited.
The sources marked with `B' and `C' also have steep spectra ($\alpha=-1.61\pm0.08$ and $\alpha=-1.78\pm0.12$ , respectively), although not as steep as `A'.
Region B shows a morphological connection to the BCG. This indicates that this source is likely a bent jet from a radio galaxy, where the plasma in the tail has aged, and that the BCG is moving through the ICM.
This movement can be the result of gas movements in an unrelaxed cluster and caused by a cluster merger event.
From the \textit{Chandra} X-ray observations, the dynamical state of clusters can be investigated. 
This can be done by means of the concentration parameter \citep[the ratio of the surface brightness within 100kpc over the surface brightness at 500 kpc,][]{santos2008} and the centroid shift between the radio and X-ray surface brightness peak \citep{poole2006}.
The concentration parameter and the centroid shift \citep[obtained from][]{cluster_states} indicate for Abell\,655 that this cluster is not dynamically relaxed \citep[see Fig. \ref{fig:cw},][]{cwperturbed}.
This could also explain why the BCG hosts a radio tail, as this indicates that the BCG is moving relative to the ICM due to the motions caused by a merger event.
The source marked with `C' does not have a clear optical counterpart, although an elliptical cluster galaxy ($z_{\text{spec}} = 0.121\pm \num{2.1e-5}$, \cite{sdss13}) is visible at the tip of the elongation, which could thus be the counterpart. This elliptical galaxy has an associated compact radio source visible in the uGMRT image in Fig. \ref{fig:gmrt_sdss}), and it is highlighted with a red circle. Therefore, classifying `C' as a fossil tailed radio source appears to be the most plausible classification, as there is no sign of a counter jet on the other side of the AGN.
Similar filament-like structures with steep spectra have been discovered in Abell S1136 \citep{2024PASA...41...50M}, where two filamentary structures are detected as originating from cluster radio galaxies, which are most likely also fossil tailed radio galaxies.

\subsection{Occurrence of diffuse emission below 30\,MHz}

A total of three pointings have been imaged so far below 30\,MHz with LOFAR. One is presented in this work, and the other two are presented in G24. 
In the combined sky coverage of these three pointings, there are 23 galaxy clusters detected in the second catalogue of SZ-detected sources with Planck \citep[PSZ2,][]{planck2016} within \SI{7}{\degree} of the pointing centre of each pointing.
In total, we found that in four out of these 23 Planck SZ-detected galaxy clusters, there is evidence of re-energised fossil plasma below 30~MHz. 
Three of the clusters with re-energised fossil plasma are presented in G24 (another object in G24 shows diffuse emission, probably re-energised fossil plasma, but it is not associated with a cluster in the PSZ2 catalogue), and one is presented here, under the assumption that the diffuse emission in region A is re-energised fossil plasma.
Although the sample is limited, we can extrapolate from this sample of galaxy clusters with decameter diffuse emission, expecting around 17 $\pm 8$\% of Planck clusters to host diffuse radio emission in the decametre band at a depth of $\sim \SI{15}{\milli\jansky}$ beam $^{-1}$.
The errors were determined assuming Poissonian noise.   
We obtained these estimates with a relatively limited sample, so they might not be a good representation of a full census. 
This limitation could introduce biases into our study, but we argue that the effect is limited.
Firstly, this observation was not targeted at a cluster with prior knowledge of diffuse, steep decametre emission. Instead, it is an observation of an unrelated source (3C196). Secondly, within the field of view of this observation, we have multiple galaxy clusters, and these objects are reflected in the statistics that we report.

It is notable that all four galaxy clusters with detected re-energised fossil plasma have a relatively low mass (\num{3.9e14}$M_\odot$ for Abell 655 and \num{2.4e14} $M_\odot$, \num{1.6e14}$M_\odot$ and \num{3.0e14} $M_\odot$ for the sources in G24) compared to the median mass of galaxy clusters detected with LOFAR \citep[\num{4.9e14} $M_\odot$][]{planck_lofar}.
Given that there are 421 PSZ2 clusters with a declination $> \SI{30}{\degree}$, we expected to detect emission corresponding to re-energised fossil plasma in $72\pm34$ sources based on the data presented in G24 and this work.
Further expansion of decametre surveys to cover the full northern sky will therefore be of high value for  understanding the nature of particle acceleration mechanisms in galaxy clusters.

Notably, we did not detect any decametre radio halos in any of the 23 galaxy clusters located in the coverage of G24 and this work.
In contrast, \cite{planck_lofar} reports the detection of 83 radio halos out of 273 PSZ2 galaxy clusters where the image allowed for a good determination of the morphology, which is an incidence rate of $30\pm 11$\% at 144~MHz.
The primary reason for the lack of detected radio halos in the decametre band is likely the relatively poor resolution of decametre observations, which limits our capability of subtracting compact objects and revealing radio halos.
In terms of sensitivity, the $1\sigma$ integrated flux density uncertainty for a 1 Mpc radio halo at the redshift of Abell 655 (z=0.129) is 192 mJy, and at HBA frequencies (tapered to the same scale as presented in this paper), it is 4.26 mJy.
In order to be equally bright in terms of signal-to-noise, this would require radio halos to have at least a spectral index of $\alpha < -2.1$.
In addition, fossil plasma from old AGN lobes can be confused as part of the radio halo, especially when the angular size of the galaxy cluster is relatively small.
However, recent work by \cite{2024A&A...688A.175O} on the detection of a radio halo in Abell~2256 shows that LOFAR has the required sensitivity to detect radio halos below 30~MHz.
Abell~2256 has a relatively low redshift ($z=0.058$) and high mass ($\text{M}_{500} = \num{6.2e14}\text{M}_\odot$), which gives rise to a larger angular size on the sky ($R_{500} = \SI{1273}{\kilo\parsec}$, corresponding to an angular size of 18 arcmin) compared to Abell~655 (7 arcmin).
Due to the strong scaling between the cluster mass and the radio halo power \citep[$P_{150~\text{MHz}} \sim M_{500}\,^{3.55\pm0.60},$][]{2023A&A...680A..30C}, we expect that galaxy clusters with a low-mass host radio halos that are too faint to be detected in the decametre band due to the relatively low sensitivity of decametre observations with LOFAR compared to LOFAR observations at higher frequencies.
Even before the LBA detection of Abell 2256, \cite{coma_halo} had reported the detection of a radio halo in the Coma galaxy cluster at 30.9~MHz. This was possible due to the low redshift of the Coma cluster ($z=0.0231$) and the high mass ($\num{7e14}M_\odot$).
This result shows that decametre observations of nearby high-mass clusters have a large potential for constraining particle re-acceleration mechanisms, as the large angular size makes subtraction of sources easier.

\section{Conclusions}

In this work, we have reported the detection of diffuse radio emission in the low-mass galaxy cluster Abell~655 within the decametre band at 15--30 ~MHz using LOFAR LBA observations. 
In addition, diffuse emission was detected using 650~MHz uGMRT observations in the cluster.
The diffuse emission has a total extent of about 700~kpc. Based on the morphology of the emission and the measured spectral indices, the diffuse emission in Abell~655 likely consists of several separate components. Below, we summarise our findings:
\begin{itemize}
    \item The uGMRT observations at 650~MHz confirm the presence of a radio halo, as previously detected by \citep{planck_lofar}. This radio halo has not been confidently detected below 144~MHz, as the low angular size of Abell 655 complicates accurate source subtraction.
    \item The LOFAR LBA and HBA observations reveal two elongated sources near the centre of the cluster. Both sources likely have an optical counterpart, which is also detected at 650~MHz. Their spectra are relatively steep ($\alpha_B = -1.61\pm0.08$ and $\alpha_C=-1.78\pm0.21$ respectively) and curved. These sources are likely fossil jets from radio galaxies.
    \item The LOFAR LBA observations also reveal a region with particularly steep spectrum emission to the east of Abell 655 (region A in Fig. \ref{fig:sixpanels}), which is completely undetected at 650~MHz. Its spectrum is ultra steep ($\alpha_A=-2.15\pm 0.24$) and curved. Such a spectral shape is expected for remnant plasma from previous AGN outbursts that have been re-energised via adiabatic compression. 
\end{itemize}

Only two radio halos have been detected in the decametre band so far \citep[in Abell 2256 and the Coma cluster,]{2024A&A...688A.175O,coma_halo}.
There are two reasons for this: Firstly, the low resolution of decametre observations prevents accurate subtraction of the point sources, which complicates the detection of the radio halo, particularly when the radio halo has a relatively small angular size when the cluster is located at a higher redshift. Other forms of diffuse radio emission in galaxy clusters, in particular ultra steep spectrum radio emission from re-energised fossil plasma, can also overpower the contribution from the radio halo.
With LOFAR 2.0, LBA and HBA observations can be performed simultaneously, which will greatly enhance the ability to subtract point sources.
Secondly, decametre observations in general have a lower sensitivity compared to higher frequency observations, which makes it harder to detect faint radio halos associated with galaxy clusters that have a lower mass, given the strong dependence of the mass on the radio halo luminosity. Therefore, observations of radio halos in the decametre band will be most promising for nearby massive systems.

Combining this work with \cite{tmpDeca}, a total of four PSZ2 clusters containing re-energised fossil plasma have been found. Based on the total number of PSZ2 clusters in the surveyed region of sky, we find that the percentage of PSZ2 clusters hosting re-energised fossil plasma is $17 \pm8 \%$.
Therefore, when we expand our surveyed region to the full northern sky above $\SI{30}{\degree}$, we should detect $72\pm34$ galaxy clusters in the PSZ2 catalogue with re-energised fossil plasma. 
Although the sample size is currently limited, this means that a much larger sample of these objects can be compiled, which in turn will help advance understanding of the connection of ancient AGN outbursts with diffuse radio emission.\\

\section*{Data availability}
The reduced LOFAR LBA images (between 15--30 MHz and 30--60 MHz), as well as the GMRT Band 4 image of Abell\,655 is available via CDS via anonymous ftp to cdsarc.u-strasbg.fr (130.79.128.5) or via http://cdsweb.u-strasbg.fr/cgi-bin/qcat?J/A+A/.

\begin{acknowledgements}
{\small CG and RJvW acknowledge support from the ERC Starting Grant ClusterWeb 804208.
AB acknowledges financial support from the European Union - Next Generation EU. 
FdG acknowledges support from the ERC Consolidator Grant ULU 101086378. 
The Dunlap Institute is funded through an endowment established by the David Dunlap family and the University of Toronto.
This paper is based (in part) on data obtained with the International LOFAR Telescope (ILT) under project code \verb+COM_LBA_SPARSE+. LOFAR \citep{lofar} is the Low Frequency Array designed and constructed by ASTRON. It has observing, data processing, and data storage facilities in several countries, that are owned by various parties (each with their own funding sources), and that are collectively operated by the ILT foundation under a joint scientific policy. The ILT resources have benefitted from the following recent major funding sources: CNRS-INSU, Observatoire de Paris and Universit\'e d'Orl\'eans, France; BMBF, MIWF-NRW, MPG, Germany; Science Foundation Ireland (SFI), Department of Business, Enterprise and Innovation (DBEI), Ireland; NWO, The Netherlands; The Science and Technology Facilities Council, UK.

We thank the staff of the GMRT that made these observations possible. GMRT is run by the National Centre for Radio Astrophysics of the Tata Institute of Fundamental Research. This research has made use of data obtained from the \textit{Chandra} Data Archive and software provided by the \textit{Chandra} X-ray Center (CXC) in the application packages CIAO and Sherpa.

Funding for SDSS-III has been provided by the Alfred P. Sloan Foundation, the Participating Institutions, the National Science Foundation, and the U.S. Department of Energy Office of Science. The SDSS-III web site is \url{http://www.sdss3.org/}. SDSS-III is managed by the Astrophysical Research Consortium for the Participating Institutions of the SDSS-III Collaboration including the University of Arizona, the Brazilian Participation Group, Brookhaven National Laboratory, Carnegie Mellon University, University of Florida, the French Participation Group, the German Participation Group, Harvard University, the Instituto de Astrofisica de Canarias, the Michigan State/Notre Dame/JINA Participation Group, Johns Hopkins University, Lawrence Berkeley National Laboratory, Max Planck Institute for Astrophysics, Max Planck Institute for Extraterrestrial Physics, New Mexico State University, New York University, Ohio State University, Pennsylvania State University, University of Portsmouth, Princeton University, the Spanish Participation Group, University of Tokyo, University of Utah, Vanderbilt University, University of Virginia, University of Washington, and Yale University. This work made use of Astropy (\url{http://www.astropy.org}): a community-developed core Python package and an ecosystem of tools and resources for astronomy \citep{astropy:2013, astropy:2018, astropy:2022}. }
\end{acknowledgements}

\begin{appendix}
\onecolumn
\section{SDSS DR9 colour image with uGMRT Band 4 radio contours}
\begin{figure*}[h!]
    \centering
    \includegraphics[width=0.8\textwidth]{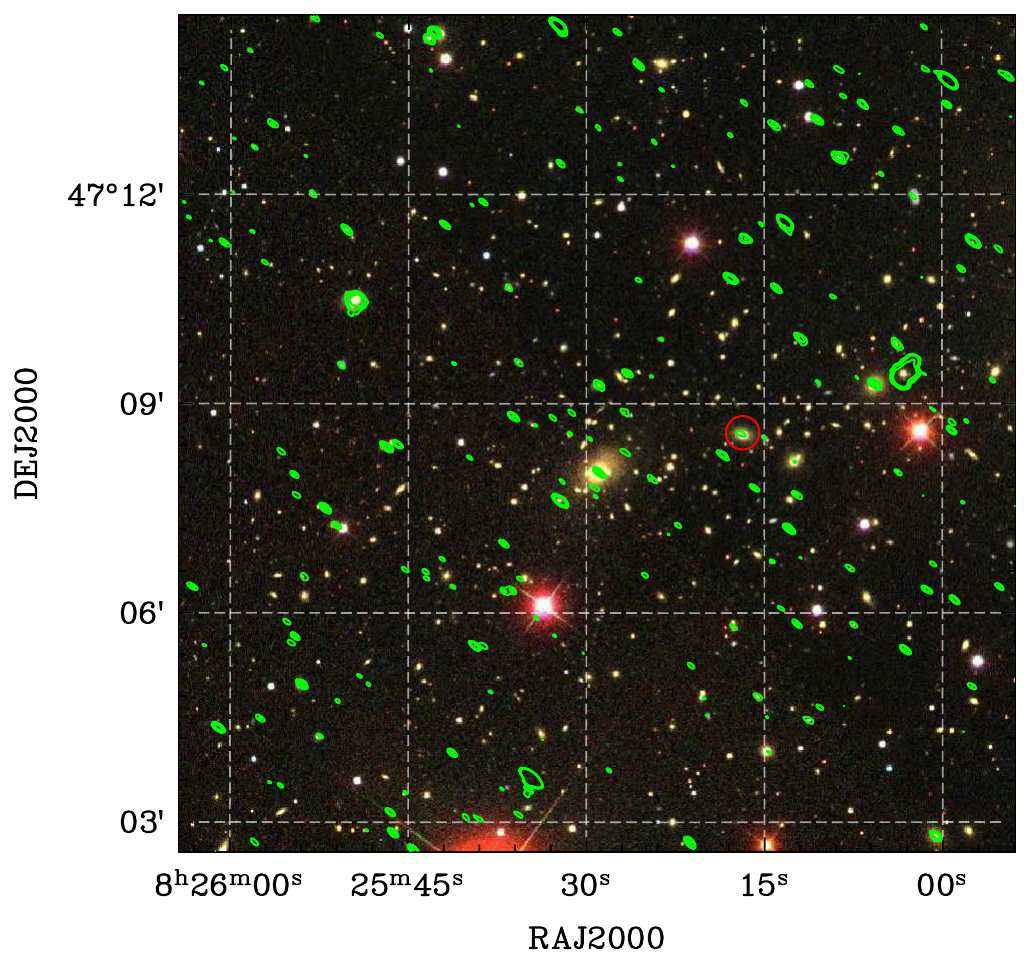}
    \caption{SDSS DR9 colour image with uGMRT Band 4 full resolution overlay. The GMRT contour data are the same as presented in the bottom-left panel in Fig. \ref{fig:sixpanels}. The contours are placed at $[4,8,16]\times \sigma$; $\sigma=\SI{11}{\micro\jansky}$. A red circle marks a detected radio source that coincides with a large elliptical cluster galaxy, which indicates that the region marked with a `C' in Fig. \ref{fig:sixpanels} is a tailed radio galaxy.}
    \label{fig:gmrt_sdss}
\end{figure*}
\FloatBarrier
\end{appendix}

\end{document}